\documentclass[12pt]{article}
\usepackage{psfrag}
\usepackage{graphics}
\usepackage{amssymb,epsfig,amsmath,euscript,array}
\usepackage{cite}

\newcounter{multieqs}

\newcommand{\be}{\begin{equation}}
\newcommand{\ee}{\end{equation}}
\newcommand{\eq}[1]{(\ref{#1})}

\newcommand{\bra}[1]{\langle #1|}
\newcommand{\ket}[1]{|#1 \rangle}

\newcommand{\bm}[1]{\mbox{\boldmath $#1$}}

\def\bd{\begin{document}}
\def\ed{\end{document}}
\def\nn{\nonumber}
\def\bea{\begin{eqnarray}}
\def\eea{\end{eqnarray}}
\let\bm=\bibitem
\let\la=\label

\def\npb#1#2#3{Nucl. Phys. {\bf{B#1}} #3 (#2)}
\def\plb#1#2#3{Phys. Lett. {\bf{#1B}} #3 (#2)}
\def\prl#1#2#3{Phys. Rev. Lett. {\bf{#1}} #3 (#2)}
\def\prd#1#2#3{Phys. Rev. {D \bf{#1}} #3 (#2)}
\def\cmp#1#2#3{Comm. Math. Phys. {\bf{#1}} #3 (#2)}
\def\cqg#1#2#3{Class. Quantum Grav. {\bf{#1}} #3 (#2)}
\def\nppsa#1#2#3{Nucl. Phys. B (Proc. Suppl.) {\bf{#1A}}#3 (#2)}
\def\ap#1#2#3{Ann. of Phys. {\bf{#1}} #3 (#2)}
\def\ijmp#1#2#3{Int. J. Mod. Phys. {\bf{A#1}} #3 (#2)}
\def\rmp#1#2#3{Rev. Mod. Phys. {\bf{#1}} #3 (#2)}
\def\mpla#1#2#3{Mod. Phys. Lett. {\bf A#1} #3 (#2)}
\def\jhep#1#2#3{J. High Energy Phys. {\bf #1} #3 (#2)}
\def\atmp#1#2#3{Adv. Theor. Math. Phys. {\bf #1} #3 (#2)}

\newcommand{\EQ}[1]{\begin{equation} #1 \end{equation}}
\newcommand{\AL}[1]{\begin{subequations}\begin{align} #1
\end{align}\end{subequations}}
\newcommand{\SP}[1]{\begin{equation}\begin{split} #1 \end{split}\end{equation}}
\newcommand{\ALAT}[2]{\begin{subequations}\begin{alignat}{#1} #2
\end{alignat}\end{subequations}}
\def\beqa{\begin{eqnarray}}
\def\eeqa{\end{eqnarray}}
\def\beq{\begin{equation}}
\def\eeq{\end{equation}}

\def\N{{\cal N}}
\def\sst{\scriptscriptstyle}
\def\thetabar{\bar\theta}
\def\Tr{{\rm Tr}}
\def\one{\mbox{1 \kern-.59em {\rm l}}}

\def\a{\alpha}      \def\da{{\dot\alpha}}
\def\b{\beta}       \def\db{{\dot\beta}}
\def\c{\gamma}  \def\C{\Gamma}  \def\cdt{\dot\gamma}
\def\d{\delta}  \def\D{\Delta}  \def\ddt{\dot\delta}
\def\e{\epsilon}        \def\vare{\varepsilon}
\def\f{\phi}    \def\F{\Phi}    \def\vvf{\f}
\def\h{\eta}
\def\k{\kappa}
\def\l{\lambda} \def\L{\Lambda}
\def\m{\mu} \def\n{\nu}
\def\o{\omega}
\def\p{\pi} \def\P{\Pi}
\def\r{\rho}
\def\s{\sigma}  \def\S{\Sigma}
\def\t{\tau}
\def\th{\theta} \def\Th{\Theta} \def\vth{\vartheta}
\def\X{\Xeta}
\def\z{\zeta}

\def\cA{{\cal A}} \def\cB{{\cal B}} \def\cC{{\cal C}}
\def\cD{{\cal D}} \def\cE{{\cal E}} \def\cF{{\cal F}}
\def\cG{{\cal G}} \def\cH{{\cal H}} \def\cI{{\cal I}}
\def\cJ{{\cal J}} \def\cK{{\cal K}} \def\cL{{\cal L}}
\def\cM{{\cal M}} \def\cN{{\cal N}} \def\cO{{\cal O}}
\def\cP{{\cal P}} \def\cQ{{\cal Q}} \def\cR{{\cal R}}
\def\cS{{\cal S}} \def\cT{{\cal T}} \def\cU{{\cal U}}
\def\cV{{\cal V}} \def\cW{{\cal W}} \def\cX{{\cal X}}
\def\cY{{\cal Y}} \def\cZ{{\cal Z}}

\def\ua{\underline{\alpha}}
\def\ub{\underline{\phantom{\alpha}}\!\!\!\beta}
\def\uc{\underline{\phantom{\alpha}}\!\!\!\gamma}
\def\um{\underline{\mu}}
\def\ud{\underline\delta}
\def\ue{\underline\epsilon}
\def\una{\underline a}\def\unA{\underline A}
\def\unb{\underline b}\def\unB{\underline B}
\def\unc{\underline c}\def\unC{\underline C}
\def\und{\underline d}\def\unD{\underline D}
\def\une{\underline e}\def\unE{\underline E}
\def\unf{\underline{\phantom{e}}\!\!\!\! f}\def\unF{\underline F}
\def\unm{\underline m}\def\unM{\underline M}
\def\unn{\underline n}\def\unN{\underline N}
\def\unp{\underline{\phantom{a}}\!\!\! p}\def\unP{\underline P}
\def\unq{\underline{\phantom{a}}\!\!\! q}
\def\unQ{\underline{\phantom{A}}\!\!\!\! Q}
\def\unH{\underline{H}}

\def\As {{A \hspace{-6.4pt} \slash}\;}
\def\bs {{b \hspace{-6.4pt} \slash}\;}
\def\Ds {{D \hspace{-6.4pt} \slash}\;}
\def\ds {{\del \hspace{-6.4pt} \slash}\;}
\def\ss {{\s \hspace{-6.4pt} \slash}\;}
\def\ks {{ k \hspace{-6.4pt} \slash}\;}
\def\ps {{p \hspace{-6.4pt} \slash}\;}
\def\pas {{{p_1} \hspace{-6.4pt} \slash}\;}
\def\pbs {{{p_2} \hspace{-6.4pt} \slash}\;}

\def\Fh{\hat{F}}
\def\Vh{\hat{V}}
\def\Xh{\hat{X}}
\def\ah{\hat{a}}
\def\xh{\hat{x}}
\def\yh{\hat{y}}
\def\ph{\hat{p}}
\def\xih{\hat{\xi}}

\def\psit{\tilde{\psi}}
\def\Psit{\tilde{\Psi}}
\def\tht{\tilde{\th}}

\def\At{\tilde{A}}
\def\Qt{\tilde{Q}}
\def\Rt{\tilde{R}}
\def\Nt{\tilde{N}}

\def\at{\tilde{a}}
\def\st{\tilde{s}}
\def\ft{\tilde{f}}
\def\pt{\tilde{p}}
\def\qt{\tilde{q}}
\def\vt{\tilde{v}}
\def\nt{\tilde{n}}

\def\delb{\bar{\partial}}
\def\bz{\bar{z}}
\def\bD{\bar{D}}
\def\bB{\bar{B}}

\def\bk{{\bf k}}
\def\bl{{\bf l}}
\def\bp{{\bf p}}
\def\bq{{\bf q}}
\def\br{{\bf r}}
\def\bx{{\bf x}}
\def\by{{\bf y}}
\def\bR{{\bf R}}
\def\bV{{\bf V}}

\def\d{\delta}\def\D{\Delta}\def\ddt{\dot\delta}

\def\pa{\partial} \def\del{\partial}
\def\xx{\times}
\def\uno{\mbox{1 \kern-.59em {\rm l}}}

\def\trp{^{\top}}
\def\inv{^{-1}}
\def\dag{{^{\dagger}}}
\def\pr{^{\prime}}

\def\rar{\rightarrow}
\def\lar{\leftarrow}
\def\lrar{\leftrightarrow}

\newcommand{\0}{\,\!}      
\def\one{1\!\!1\,\,}
\def\im{\imath}
\def\jm{\jmath}

\newcommand{\tr}{\mbox{tr}}
\newcommand{\slsh}[1]{/ \!\!\!\! #1}

\def\vac{|0\rangle}
\def\lvac{\langle 0|}

\def\hlf{\frac{1}{2}}
\def\ove#1{\frac{1}{#1}}

\def\Box{\square}
\def\ZZ{\mathbb{Z}}
\def\CC#1{({\bf #1})}
\def\bcomment#1{}
\def\bfhat#1{{\bf \hat{#1}}}
\def\VEV#1{\left\langle #1\right\rangle}

\newcommand{\ex}[1]{{\rm e}^{#1}} \def\ii{{\rm i}}

\begin{document}

\hfill{hep-th/0206167}

\vspace{20pt}

\begin{center}

{\Large \bf PP-wave string interactions from \\}
\vspace{10pt}
{\Large \bf $n$-point correlators of
BMN operators  }

\vspace{30pt}

{\bf Chong-Sun Chu, Valentin V. Khoze
and Gabriele Travaglini}

{\small \em Centre for Particle Theory,
University of Durham, Durham, DH1 3LE, UK}

\vspace{10pt}

Email: {\sffamily \tt chong-sun.chu, valya.khoze,
gabriele.travaglini@durham.ac.uk }

\vspace{30pt}
{\bf Abstract}

\end{center}

BMN operators are characterized by the fact that they have infinite
$R$-charge and finite anomalous dimension in the BMN double scaling
limit. Using this fact,
we show that the BMN operators close under operator product
expansion and form a sector in the $\cN=4$ supersymmetric Yang-Mills theory.
We then identify short-distance limits of general BMN
$n$-point correlators, and show how they correspond to the pp-wave
string interactions.
We also discuss instantons in the light of the pp-wave/SYM
correspondence.

\vspace{0.5cm}

\setcounter{page}0
\thispagestyle{empty}
\newpage


\section{Introduction}

Recently Berenstein, Maldacena and Nastase (BMN) \cite{BMN}
put forward a remarkable proposal of a correspondence between certain
operators in  $\cN=4$ supersymmetric $SU(N)$ Yang-Mills theory (SYM) and massive
states in string theory in a pp-wave background \cite{bfhp}
\begin{equation} \label{bmn1}
\frac{1}{\sqrt{J} N^{J/2+1}} \Tr Z^J  \longleftrightarrow \ket{0,p^+},
\end{equation}
\begin{equation}\label{bmn2}
\frac{1}{\sqrt{J} N^{J/2+1}} \sum_{l=0}^J \Tr [\phi_3 Z^l \phi_4
Z^{J-l}] e^{\frac{2 \pi i n l}{J} } \longleftrightarrow a^{7 \dag}_n
a^{8 \dag}_{-n} \ket{0,p^+}  .
\end{equation}
It is often said that the BMN operators
form a {\it sector} in SYM
in the double scaling limit
\begin{equation} \label{double}
N \to \infty\ , \quad J \sim \sqrt{N} \quad
\,\mbox{with\, $g_{\rm YM}$\, fixed}.
\end{equation}
In this paper, we will give a more precise
meaning to this statement using the operator product expansion (OPE).
We will argue that, in the double scaling limit,
the OPE of  BMN operators does not give
rise to non-BMN operators.
We will use this {\it short OPE} to analyse certain short distance
limits of general $n$-point correlators of the BMN operators, and
find a precise correspondence with the structure of the string
interactions
in the pp-wave background.

The BMN correspondence  holds in
the double scaling limit \eq{double}.
In this limit the \mbox{'t Hooft}
coupling $\lambda=g_{\rm YM}^2 N$ is infinite and perturbative
calculations in gauge theory are not under control.
BMN instead concentrated on  a class of `near-BPS' operators
with large $R$-charge $J$, e.g. as in \eq{bmn2}.
For these operators the coupling is effectively
\begin{equation} \label{lampr}
 \l' = \frac{g_{\rm YM}^2 N}{J^2} = \frac{1}{(\mu p^+ \a')^2}\ ,
\end{equation}
which is  finite in the large $N$ limit \eq{double} and can be
taken small. The  anomalous dimensions $\d$ of the BMN operators
are finite in the limit \eq{double}, and are related to the masses
of the corresponding string states via\footnote{The BMN operator
on the LHS of \eq{bmn2} has two insertions of $\phi$ fields and
its anomalous
dimension is
$\delta=\Delta-J-2$.
We will adopt the complex scalar field notation
$Z=\phi_1+i\phi_2$, $\Phi=\phi_3+i\phi_4$ and
$\Psi=\phi_5+i\phi_6$. ``Nonholomorphic'' BMN operators, e.g. of
the form
$\sum_l {\rm Tr} [\Phi Z^l \bar{\Psi} Z^{J-l}] e^{\frac{2 \pi i
nl}{J} }$, are also allowed. What is not allowed are the operators
with insertions of $\Phi$ and $\bar{\Phi}$ or $\Psi$ and
$\bar{\Psi}$ at the same time.}
\begin{equation} \label{anomdim}
\Delta -J= H_{\rm lc}/\mu,
\end{equation}
where $H_{\rm lc}$ is the lightcone string Hamiltonian and $\mu$ is the
scale of the pp-wave metric. Written in terms of gauge theory
parameters, the string theory gives a  prediction for the conformal
dimension of the BMN operators
\begin{equation} \label{dJ}
(\Delta -J)_n = \sqrt{1+ \frac{g_{\rm YM}^2 N n^2}{J^2}}.
\end{equation}
Since all the  states in the perturbative spectrum of the string
theory  are  already accounted for by the full set of BMN operators,
it is a central part of the BMN proposal that the
anomalous dimensions
of the other (non-BMN) operators become infinite\footnote{
Indeed it appears so in perturbation theory.
}
in this limit
\eq{double}, and hence  the non-BMN operators
play no r{\^o}le in the perturbative pp-wave/SYM correspondence.

The field theory side of the pp-wave/SYM
correspondence was recently discussed in \cite{seme,dzf,gross,CKT}.
Non-planar
diagrams in the BMN limit \eq{double} were first studied in \cite{seme} and in
\cite{dzf} and were found to be important and governed by
$J^4/N^2$.  It follows from the double scaling limit \eqref{double}
that in addition to $\l'$ defined in \eqref{lampr}, there is a
second dimensionless constant
\begin{equation}
g_2 :=\frac{J^2}{N}= 4 \pi g_s (\mu p^+ \a')^2 \ ,
\end{equation}
which plays the r{\^o}le of the genus counting parameter for the SYM
Feynman diagrams \cite{seme,dzf}.
Anomalous dimensions were computed in \cite{seme,dzf,gross,zanon}.
It was proposed in \cite{dzf} that the coefficient of the three-point
function of BMN operators in SYM
is related to the three-string interactions in the pp-wave background.
Planar three-point functions of BMN operators in free
field theory were calculated in \cite{dzf} and in the first nontrivial
order of $\l'$ in \cite{CKT}. The proposal of \cite{dzf} states that
the matrix element of the lightcone Hamiltonian is related to the
coefficient $C_{ijk}$ of the three-point function in field theory via
\begin{equation}
\bra{i} P^- \ket{j,k} = \mu (\D_i -\D_j -\D_k) C_{ijk}
\end{equation}
in the leading order in $\l'$. Checks of this in the free field limit
were performed in \cite{dzf,sv,lee}.
See \cite{gopakumar,ver}
for further aspects about string interactions in pp-wave
background.
Another form of this proposal (which is insensitive to the prefactor)
\cite{huang} relates the ratio of the
three-string interactions
with those of the field theory three-point function coefficients
\begin{equation} \label{hh}
\frac{ \langle \Phi_1| \langle \Phi_2|\langle \Phi_3| V \rangle }
{\langle 0_1| \langle 0_2|\langle 0_3| V\rangle }=
\frac{ C_{123} }{ C^{(\rm vac)}_{123} }.
\end{equation}
Here the left hand side is the normalized 3-string interaction
in the string field theory  formalism \cite{sv},
and  $\ket{V}$ is the lightcone three-string vertex.
In \cite{CKT} the field theory results for the three-point function
and the corresponding string theory prediction
were derived and found to be in  precise agreement,
thus confirming \eq{hh} up to and including the order $\l'$ corrections.

The next important problem to understand is how the higher point
SYM correlation functions manifest themselves on the string theory
side of the correspondence. This issue will be addressed in the
present paper. The plan of the paper is as follow. In section 2,
we show that the OPE of the BMN operators is closed in the double
scaling limit. We also show that this short OPE has a very natural
interpretation in string theory. Using this short OPE,  we
establish and extend in section 3 the correspondence between SYM
correlators and pp-wave string interactions.
We show that in a certain short distance limit involving a hierarchy of
multi-pinchings,
generic BMN correlators reduce
to expressions written in terms of
the three-point function
coefficients and the anomalous dimensions. These expressions
have a form that corresponds precisely to
tree level string interactions in pp-wave background\footnote{
Very recently, after the first version of this paper appeared, it
was argued in \cite{new1} that four- and higher-point functions
are ill-defined in the BMN limit \eqref{double}. We will point out
in section 3 that our approach based on the OPE requires a
specific order of limits: first we take the short-distance
(pinching) limit, and then we take the BMN limit. In this case it
follows that the inconsistencies mentioned in sections 2.2 and 2.3
of \cite{new1} are not present.}.
We also briefly discuss how 
loop corrected string interactions
can be extracted  from the BMN correlators.
In section 4, we analyze instanton contributions to two- and
three-point functions of BMN operators. We use a simple argument
based on counting of fermion zero modes in the instanton
background to show that two-point functions are protected from
instanton corrections. This is consistent with the apparent
absence of D-instanton solutions in pp-wave background. However,
for generic three- and higher-point BMN correlators, our simple
argument is not sufficient since fermion zero modes can be
saturated in this case and one would require a detailed
calculation to determine if the instanton corrections to the
three-point functions are present.

Other relevant aspects of the correspondence have been studied in
\cite{s1}, where  it was emphasised that the worldsheet model
is exactly solvable in the lightcone gauge.
Questions of holographic relation in the pp-wave context were considered in
\cite{hr}.

\section{Short OPE of BMN operators}

We first briefly recall the structure of operator product
expansion (OPE). In a general QFT, the OPE is the statement that, in
the short distance limit, the
product of two local operators
can be expressed in terms of a sum over
local operators in the theory
\begin{equation}\label{OPE}
\cO_I(0) \cO_J(x)= \sum_K c_{IJ}^K(x) \cO_K(0).
\end{equation}
This is an operator relation. For  generic correlators $\langle
\cO_I(0) \cO_J(x)  \prod_i A_i(y_i) \rangle $, the relation
\eq{OPE} holds only when $|x| \ll |y_i|$. For conformally
invariant unitary theories, one can always choose to work with a
basis of operators which do not mix with each other and have
definite conformal dimensions.
If the operators are also conformal primary operators, then
conformal invariance  of the theory implies that the two-point and
three-point functions can be written in the  form
\begin{equation} \label{2pt}
\langle {\cO}_I (x_1) \cO_J(x_2) \rangle = \frac{\d_{IJ}}{(4 \pi^2
x_{12}^2)^{\Delta_I}},
\end{equation}
\begin{equation} \label{3pt}
\langle \cO_{I_1}(x_1) \cO_{I_2}(x_2) \cO_{I_3}(x_3) \rangle  =
\frac{C_{{I_1} {I_2} {I_3}} }
{(4\pi^2 x_{12}^2)^{\frac{\D_1+\D_2 -\D_3}{2}}
 (4\pi^2 x_{13}^2)^{\frac{\D_1+\D_3 -\D_2}{2}}
 (4\pi^2 x_{23}^2)^{\frac{\D_2+\D_3 -\D_1}{2}}},
\end{equation}
where $x_{IJ}^2: = (x_I-x_J)^2$.
The OPE takes then a simple form
\begin{equation}\label{OPE1}
\cO_{I}(0)\cO_{J}(x) = \sum_K \frac{C_{IJK}}{|x|^{\Delta_I
+\Delta_J -\Delta_K}} \; \cO_{K}(0).
\end{equation}

We will now consider the OPE of two BMN operators. It is known
\cite{seme,dzf} that the original single-trace BMN operators (for
example, the LHS of \eq{bmn2}) mix at the nonplanar level among
themselves and with multi-trace operators
\cite{bianchi2,new1,new2}.
Hence the original BMN operators do not
have well defined conformal dimensions, and one has to define a
new basis where the operators do not mix \cite{seme,new1}. This
redefinition has to be implemented to all orders in $\l'$ and
$g_2$. Two different choices of these re-diagonalized bases were
recently considered in \cite{new1,new3}.

At present we don't know why (and if) the relevant SYM operators
should be conformal primaries. However it has been shown recently
that the three-point functions constructed using the
re-diagonalized basis of BMN operators derived in \cite{new1} take
the form \eqref{3pt} (at least to the first non-trivial order in
$\l'$). In what follows, we will restrict ourselves to BMN
operators with scalar impurities only and always assume the
re-diagonalized basis of \cite{new1}, so that \eqref{OPE1} holds.

As we mentioned earlier,
anomalous dimensions $\d_K$ of non-BMN
operators become infinite in the double scaling limit, and hence they
do not appear in the OPE. As a result, the sum in \eq{OPE1}
is reduced  to BMN operators only,
\begin{equation}\label{sOPE}
\cO_{1}(0)\cO_{2}(x) =
\frac{1}{|x|^{ \Delta_1 + \Delta_2 - J }}
\sum_{K \in {\rm BMN}} C_{12K} \; |x|^{\Delta_K- J} \; \cO_{K}(0), \quad x \to 0.
\end{equation}
Here $J= |J_1 \pm J_2|$ depending on whether $\cO_1$ and $\cO_2$
have the same or opposite sign of R-charge. Note that $\Delta_K-
J= n_K + \delta_K$, where $n_K$ is the total engineering dimension
of the impurities \footnote{Each scalar impurity contributes 1;
each fermion impurity contributes $3/2$ and each derivative
impurity contributes 2.}  in the operator $\cO_{K}$  and  $\d_K$
is the anomalous dimension.
In the double scaling limit $\d_K$ is finite only for the BMN
operators, hence only BMN operators contribute to the sum in
\eqref{sOPE} since $x$ is small.


To make the statement \eq{sOPE} more precise, let us
consider a general $n$-point function $\langle \cO_1(0)\cO_2 (x) \prod_i A_i(y_i)
\rangle$ where $\cO_1$ and $\cO_2$ are BMN operators, and $A_i$ are some local
operators. Using
\eq{OPE1} with fixed $|y_i| \gg |x|$,   we obtain
\bea \label{decouple}
\langle \cO_1(0)\cO_2 (x)  \prod_i A_i(y_i) \rangle &=&
\sum_K \frac{C_{12K}}
{|x|^{\Delta_1 +\Delta_2 -\Delta_K}} \;
\langle\cO_{K}(0) \prod_i A_i(y_i) \rangle \nn\\
&=& \frac{1}{|x|^{\Delta_1 +\Delta_2 - J}}
\sum_{K\in {\rm BMN}} C_{12K} \left|\frac{x}{y_1}\right|^{\Delta_K -J}
f_K(y_i)
\eea
where we have defined $ \langle\cO_{K}(0) \prod_i A_i(y_i) \rangle :=
|y_1|^{-(\D_K -J)} f_K(y_i)$.
In the last line of \eq{decouple}, we have
used $|x/y_1| \ll 1$ and
the fact that $\D_K -J \to \infty$ unless $\cO_K$ is a BMN operator. This
demonstrates the shortening of the OPE of the BMN operators in the
double scaling limit.

The short OPE has a natural interpretation in string theory. To see
this, let us use \eq{anomdim} and rewrite \eq{sOPE} in the form
\begin{equation}\label{limit-form}
|x|^{\D_1 +\D_2 - J } \cO_1(0) \cO_2(x) =
\sum_{K\in {\rm BMN}}\;  C_{12K}\; |x|^{E_K/\mu}\; \cO_K(0), \quad x\to 0.
\end{equation}
Here we have introduced the notation ${E_K/\mu}:=\Delta_K- J=
n_K+\delta_K$, cf. Eq. \eqref{anomdim}. The factor $C_{12K}$
corresponds to the 3-string interaction vertex \cite{dzf}, see
\eq{hh}.
The factor $|x|^{E_K/\mu}$ corresponds to,
apart from an overall measure factor,
the integrand of the string propagator (without ghosts)
\begin{equation}
\frac{1}{L_0+ \bar{L}_0 -2} = \int \frac{d^2 q}{|q|^2} \; q^{L_0 -1} \;
\bar{q}^{\bar{L}_0-1},
\end{equation}
with modulus $ |q|^2$ mapped to $|x|^{1/\mu}$. Equation
\eq{limit-form} has a suggestive diagrammatic representation, figure 1.
We remark that the correspondence in figure 1 relies on
the fact that non-BMN operators do not appear in the OPE \eq{limit-form}.
We will
use this fundamental relation to uncover the higher point string
interactions from the short distance limits of the BMN correlators.

\begin{figure}[ht]
\label{fig1}
\psfrag{o1}{$\cO_1$}
\psfrag{o2}{$\cO_2$}
\psfrag{o3}{$\sum_K C_{12K} \cO_K$}
\psfrag{x2}{$x_2$}
\psfrag{s1}{$\ket{1}$}
\psfrag{s2}{$\ket{2}$}
\psfrag{sk}{$\ket{K}$}
\psfrag{eq}{$\longleftrightarrow$}
\begin{center}
{\scalebox{1}{
\includegraphics{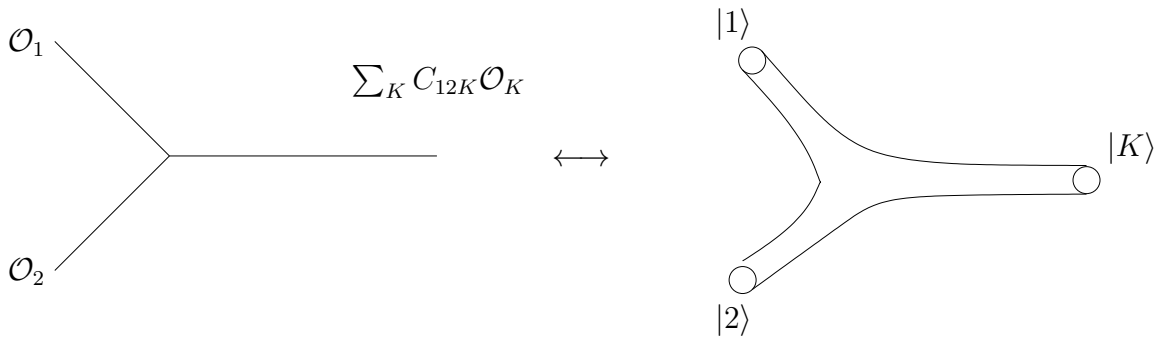}}
}
\end{center}
\caption{Short OPE of BMN operators and its string interpretation. }
\end{figure}

\section{$n$-string interactions from BMN correlators}

In this section, we study correlators of  BMN operators and
their relation to pp-wave string interactions.
For each string interaction process, we will divide the set of
states into incoming and outgoing,
according to whether they have  negative or positive  values of $p^+$.
We take the convention that incoming states are associated with
BMN operators made out of $\bar{Z}$
and outgoing states are associated with BMN operators made out of $Z$.
In the analysis below, we will denote incoming operator by
$\bar{\cO}$ and outgoing operator by  ${\cO}$.

It is well known that
the form of three-point functions in $\cN=4$ SYM is uniquely
determined by conformal invariance. Hence, it is natural to expect
that the $x$-independent coefficient $C_{I_1 I_2 I_3}$ of the three-point
function is directly related to the three-string interaction
\cite{GS2},
describing the joining and splitting of closed strings.
The analysis carried out in \cite{dzf} and more recently in \cite{CKT}
confirms that this is indeed the case.
On the other hand,
general $n$-point functions ($n>3$)
have a non-trivial space-time dependence and their
form is not fully determined by conformal invariance.
A question then arises of what is the meaning of these
$n$-point functions of BMN operators on the string theory side.
In this section we will argue that the short OPE introduced in
the last section
leads to a natural correspondence between
the short distance limits of multi-BMN correlators and higher
string interactions.

\begin{figure}[ht]
\label{fig2}
\psfrag{1}{$1$}
\psfrag{2}{$2$}
\psfrag{3}{$3$}
\psfrag{4}{$4$}
\psfrag{k}{$K$}
\begin{center}
{\scalebox{1}{
\includegraphics{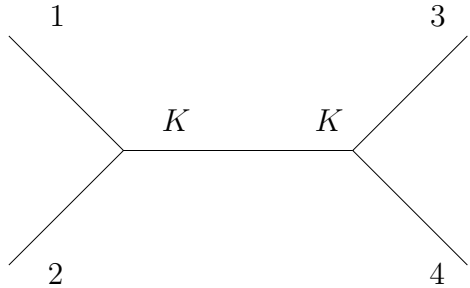}}
}
\end{center}
\caption{A four-string interaction.}
\end{figure}

We first consider a four-string
process $1+2 \to 3+4$. This corresponds to
the four-point correlation function
$\langle \bar{\cO}_1(x_1)\bar{\cO}_2(x_2)\cO_3(x_3)\cO_4(x_4)\rangle$
of BMN operators.
The s-channel of the string process corresponds to a specific double OPE
of this correlator,
such that $x_{12}\to 0$ and $x_{34}\to 0$,
as depicted in figure 2.
More precisely, consider the following expression
\bea \label{dopes}
&&|x_{12}|^{\D_1+\D_2 -J_s}|x_{34}|^{\D_3+\D_4 -J_s }
|x_{23}|^{2 J_s }
\langle \bar{\cO}_1(x_1)\bar{\cO}_2(x_2)\cO_3(x_3)\cO_4(x_4)\rangle \nn\\
&&=
\sum_{K\in {\rm BMN}}\; C_{\bar{1}\bar{2}\bar{K}} \;C_{34K}
\left|\frac{x_{12}x_{34}}{x_{23}^2}\right|^{E_K/\mu}, \quad
J_s:= J_1+J_2 =J_3+J_4 .
\eea
Here we took
a specific
{\it double pinching limit} of a conformally invariant expression
involving the four-point correlator of BMN operators. The choice of
pinching determines which operators we are expanding in the short OPE.
Hence the RHS of \eq{dopes} is obtained from the short OPE of
$\bar{\cO}_1(x_1)\bar{\cO}_2(x_2)$
and of $\cO_3(x_3)\cO_4(x_4)$. The skeleton diagram
of this double OPE is shown in figure 2 and it corresponds
to the s-channel of the string
process.
In the expression above,
\begin{equation}
J_s = J_1+J_2 =J_3+J_4
\end{equation}
is the conserved R charge flowing through the s-channel.
As before,  $C_{\bar{1}\bar{2}\bar{K}}$ and $C_{34K}$
correspond to two 3-string interaction  vertices,
and\footnote{Note that
$|x_{12}||x_{34}|/|x_{23}|^2$ is a conformally invariant cross-ratio
in the double pinching limit.}
$(|x_{12}||x_{34}|/|x_{23}|^2)^{\d_K}$ corresponds,
apart from an overall measure factor,
to the  integrand of the string propagator.
The double pinching limit in \eq{dopes} is understood as
a power-series expansion in the small quantity  $|x_{12}||x_{34}|/|x_{23}|^2$.
In other words,
only finite values of $\d_K$ are left in the sum.
The sum over the BMN operators
in \eq{dopes} corresponds precisely to the sum over {\it physical}
intermediate string states in the s-channel.

The two other channels of the four-point string
process  similarly arise from the remaining two double pinching limits
of the four-point BMN correlator:
\begin{equation}\label{dopetu-t}
\mbox{t-channel} : \;
\lim_{x_{13},x_{24}\to 0}
|x_{13}|^{\D_1+\D_3 - J_t}
|x_{24}|^{\D_2+\D_4- J_t } |x_{12}|^{2 J_t}
\times \langle
\bar{\cO}_1(x_1)\bar{\cO}_2(x_2) \cO_3(x_3)\cO_4(x_4)\rangle
\end{equation}
\begin{equation} \label{dopetu-u}
\mbox{u-channel} : \;
\lim_{ x_{14}, x_{23}\to 0}
|x_{14}|^{\D_1+\D_4 - J_u}
|x_{23}|^{\D_2+\D_3 - J_u} |x_{12}|^{2 J_u}
\times \langle
\bar{\cO}_1(x_1)\bar{\cO}_2(x_2)
\cO_3(x_3)\cO_4(x_4)\rangle .
\end{equation}
Here
\begin{equation}
J_t := |J_1 -J_3| = |J_2 -J_4| \quad\mbox{and}\quad
J_u := |J_1 -J_4| = |J_2 -J_3|.
\end{equation}
The double pinching limits in the equations above are to be interpreted as
before.

A general comment is in order. To use the OPE we have to assume
the short-distance pinching limit first. After this we take the
double-scaling limit \eqref{double} and this restricts the
operators appearing on the RHS of the OPE to the BMN operators. In
section 2.3 of \cite{new1} it was argued that the radiative
corrections to the four-point correlators become infinite when the
BMN limit is taken {\it before} the pinching limit. This does not
apply to our case. In addition, the discontinuous values of the
four-point functions reported in section 2.2 of \cite{new1}
precisely correspond to the three different channels of the
four-string process.

So far we have been concerned with reproducing the integrand of
the string
interaction in different channels. To get the full string interaction,
one has to integrate over the string moduli and sum
over the different channels in string field theory.
Our method, based on
the analysis of short distance limits of BMN correlators, does not
give information about the integration region over the string moduli.
It would be interesting to understand better how and if
this arises from the SYM point of view.

\begin{figure}[ht]
\label{fig3}
\psfrag{1}{$1$}
\psfrag{2}{$2$}
\psfrag{3}{$3$}
\psfrag{4}{$4$}
\psfrag{5}{$5$}
\psfrag{6}{$6$}
\psfrag{7}{$7$}
\psfrag{k1}{$K_1$}
\psfrag{k2}{$K_2$}
\psfrag{k3}{$K_3$}
\psfrag{L}{$L$}
\begin{center}
{\scalebox{1}{
\includegraphics{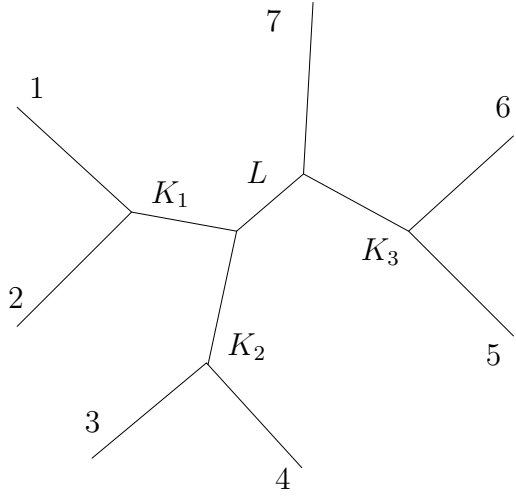}}
}
\end{center}
\caption{A seven-string interaction.}
\end{figure}

We now discuss
the generalization to higher-point string
interaction. The general
approach is similar to the analysis above, but there is an important
novel feature -- a  {\it hierarchy of pinchings}.
To illustrate this we consider an example of seven-point string
interaction
$1+2+3+4 \to 5+6 +7$
as depicted in figure 3.
As it should be clear from the structure of this string
interaction,
the corresponding field theory correlator should be taken in the
short-distance limit involving the pinching of
$x_{12}$, $x_{34}$, $x_{56}\to 0$. This has to be followed by a
second pinching of $x_{13}$, $x_{57}\to 0$. More specifically we consider
\bea
&&|x_{12}|^{\D_1+\D_2 -J_{12}}
|x_{34}|^{\D_3+\D_4 - J_{34}}
|x_{56}|^{\D_5+\D_6 - J_{56}}
|x_{57}|^{\D_7-J_7}
|x_{15}|^{2 J}
\; \langle \prod_{I=1}^4 \bar{\cO}_I (x_I)
\prod_{I=5}^7 \cO_I (x_I)
\rangle  \nn \\
&&\stackrel{\mbox{\sl (1)}}{\longrightarrow}
|x_{57}|^{E_7/\mu}
|x_{15}|^{2 J}
\sum_{K_1,K_2,K_3} C_{\bar{1}\bar{2}\bar{K_1}}C_{\bar{3}\bar{4}\bar{K_2}}
C_{56K_3}
|x_{12}|^{E_{K_1}/\mu}|x_{34}|^{E_{K_2}/\mu}|x_{56}|^{E_{K_3}/\mu } \\
&&\hspace{180pt}
\cdot \langle
\bar{\cO}_{K_1}(x_1)\bar{\cO}_{K_2}(x_3) \cO_{K_3}(x_5)\cO_{7}(x_7)
\rangle \nn \\
&&\stackrel{\mbox{\sl (2)}}{\longrightarrow}
\sum_{K_1,K_2,K_3,L}
C_{\bar{1}\bar{2}\bar{K_1}}C_{\bar{3}\bar{4}\bar{K_2}}
C_{56K_3}
C_{\bar{K_1} \bar{K_2} \bar{L}}C_{K_3 7L}
\left| \frac{x_{12}}{x_{13}}\right|^{E_{K_1}/\mu}
\left| \frac{x_{34}}{x_{13}}\right|^{E_{K_2}/\mu}
\left| \frac{x_{56}}{x_{57}}\right|^{E_{K_3}/\mu}
\left| \frac{x_{13}x_{57}}{x_{15}^2}\right|^{E_{L}/\mu} , \nn
\eea
where the limit {\sl (1)} denotes the first hierarchical pinching
$|x_{12}|,\;|x_{34}|,\;|x_{56}| \to 0$, and the limit {\sl (2)} denotes the
second pinching $|x_{13}|,\;|x_{57}|\to 0$.
In the above expression,
\begin{equation}
J_{12} := J_1 +J_2 \quad\mbox{etc},
\end{equation}
and $J$ is
the total R charge
\begin{equation}
J := J_1 + J_2 +J_3 +J_4
\end{equation}
As before, the three-point coefficients $C_{IJK}$ correspond to the
three-string vertices, and the $x$-ratios are identified with the moduli
of the corresponding string propagators.
The sum over all such inequivalent skeleton diagrams in field theory,
appropriately integrated over, corresponds to the full string interaction.

In the above, we have discussed how, in a certain short distance
limit (which involves a hierarchy of multi-pinchings) an $n$-point BMN
correlator corresponds to a tree $n$-string interaction
in a specific channel.  Now we discuss how to obtain string loop
interactions
from the  BMN correlators.  The analysis again hinges on the use of the short
OPE, but there is yet another important new feature of the procedure --
cluster decomposition. To illustrate the idea,
it is sufficient to consider the simple case of a one-loop two-point string
interaction
drawn in figure 4. To reproduce this two-point amplitude, we have
to consider a {\it six}-point BMN correlator.
\begin{figure}[ht]
\label{fig4}
\psfrag{1}{$1$}
\psfrag{2}{$2$}
\psfrag{r}{$r$}
\psfrag{s}{$s$}
\begin{center}
{\scalebox{1}{
\includegraphics{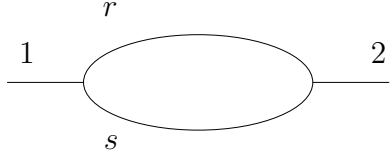}}
}
\end{center}
\caption{A one-loop corrected string interaction.}
\end{figure}

Consider the following  quantity $I$ and its limits
\bea
I:= &&\frac{|x_{12}|^{2 \D_1} |x_{46}|^{2 \D_2}}{x_{23}^{\D_1} x_{45}^{\D_2}}
\nn \\
&&\times    \sum_{r,s \in {\rm BMN}}
\langle
\cO_1(x_1) \bar{\cO}_r(x_2)
\bar{\cO}_s(x_3)\cO_r(x_4)\cO_s(x_5)
\bar{\cO}_2(x_6)\rangle
|x_{23}|^{2 \D_r + \D_s - J_r} |x_{45}|^{2 \D_s+  \D_r -J_s} \nn
\eea
\bea \label{1loop}
&&\stackrel{\mbox{\sl (1)}}{\longrightarrow} \;
\frac{|x_{12}|^{2 \D_1} |x_{46}|^{2 \D_2}}{x_{23}^{\D_1} x_{45}^{\D_2}}
\sum_{r,s,a,b \in {\rm BMN}}
C_{\bar{r}\bar{s}\bar{a}} C_{rsb}  |x_{23}|^{E_r/\mu + \D_a }
|x_{45}|^{E_s/\mu + \D_b }
\langle
\cO_1(x_1) \bar{\cO}_a(x_2) \cO_b(x_4)
\bar{\cO}_2(x_6) \rangle \;
\nn\\
&&\stackrel{\mbox{\sl (2)}}{\longrightarrow} \sum_{r,s \in {\rm
BMN}} C_{\bar{r}\bar{s}\bar{1}} C_{rs2} \; |x_{23} |^{E_r/\mu}
|x_{45} |^{E_s/\mu} \eea Here the limit {\sl (1)} denotes the
short distance limit $|x_{23}| \to 0$, $|x_{45}|  \to 0$. After
this limit, $I$ becomes a function of $x_1, x_2$ and $x_4, x_6$.
The second limit {\sl (2)} is a {\it large distance cluster limit}
where we group $x_1, x_2$ and $x_4, x_6$ into two independent
clusters and send them far away from each other. Due to the
cluster decomposition principle, which holds in a general QFT, the
four-point function in the second line of \eq{1loop} factorizes
in this limit as
\begin{equation}
\langle
\cO_1(x_1) \bar{\cO}_a(x_2) \cO_b(x_4)\bar{\cO}_2(x_6)  \rangle =
\frac{\d_{1a}}{|x_{12}|^{2\D_1}} \frac{\d_{2b}}{|x_{46}|^{2\D_2}},
\end{equation}
and the last line in \eq{1loop} follows. As before,
the three-point coefficients $C_{IJK}$ correspond to the
three-string vertices. $x_{23}$ and $x_{45}$  are identified
with the moduli of the corresponding string propagators. Note that
only physical degree of freedom propagate in the string loop in
the lightcone gauge. The last line of \eq{1loop}, when
appropriately integrated over, corresponds to the full one-loop
two-point string interaction.

It should be clear from the above analysis how to generalize to the higher
loop case.  Generally, to obtain an $n$-point $h$-loop string
interaction
for a specific string field theory diagram, one has to start with
a $3v$-point BMN correlator where  $v$ is the number of vertices in
the string  diagram. Short distance limit (like   {\sl (1)}
above) generates the string propagators. Then it is followed by a large
distance clustering limit which separates the $v$ vertices.  The resulting
expression is in direct correspondence with the string loop
interaction.

Note that in the pinching limit we proposed above, a general
string interaction is effectively reduced
 to {\it two-point correlators}
in field theory. This appears to be in agreement with a recent
proposal of Verlinde \cite{ver} which states that pp-wave string
interactions should be extracted from two-point correlators in
field theory.

\section{On instanton corrections to BMN correlators}

Due to the very nature of the Penrose limit, which relies on the
existence of null geodesics, the pp-wave metric cannot be
Euclideanized and stay real. This suggests that there are no
D-instantons in a pp-wave background.
For the pp-wave/SYM correspondence to hold, this means that there
should be no Yang-Mills instanton corrections to the SYM
quantities which are relevant for the correspondence. A priori,
there is no reason to expect that instantons do not contribute to
generic SYM correlators since $g^2_{\rm YM}$ is fixed in the
double scaling limit \eq{double}.

We  start by examining two-point functions. Our analysis
is similar to that in   \cite{bianchi}, which
showed that  instanton
corrections to extremal correlators in $\cN=4$ SYM vanish.
Recall that in the case of $\cN=4$ SYM with gauge group $SU(N)$,
the Dirac operator in the adjoint representation in the background
of an instanton of winding number $k$ has  $8kN$ zero modes.
However, only 16 of them are exact zero modes, since the remaining
ones are lifted by the presence of a fermion quadrilinear,
which is induced by the Yukawa term in the instanton action \cite{valyads}.
All the considerations in this section apply to instantons of arbitrary
charge. For full details of the ADHM
instanton calculus we refer the reader to the review \cite{icalc}.
The exact zero modes can be  generated acting with supersymmetry
and superconformal transformation on the instanton, and take the form
\begin{equation}\label{zeromodes}
\lambda^{I}(x) = \frac{1}{2} F_{\mu \nu} \sigma^{\mu \nu} \zeta^{I}(x)
\ .
\end{equation}
Here $\zeta^{I}(x) = \xi^{I} + \bar{\eta}^{I}
\bar{\sigma}_{\mu}(x - x_0)^{\mu}$ and $I=(0, i)$, with $i=1,2,3$.
$\xi^{I}$, $\bar{\eta}^{I}$ are eight constant Weyl spinors,
and $x_0$ is the centre of the instanton configuration.
Solving the equations of motion for the three complex scalar fields
$\phi^{i}$ (and $\bar{\phi}_i$)  of $\cN=4$ SYM,
 we obtain
\bea \label{phiinst}
\phi^{i}_{\rm cl} &= & \frac{1}{2}\,
\zeta^{0} F_{\mu \nu} \sigma^{\mu \nu}\zeta^{i}
\ , \\
\phi_{i\,\rm cl}^{\dagger}&= &\epsilon_{ijk} \,\zeta^{j} F_{\mu
\nu} \sigma^{\mu \nu}\zeta^{k} \  . \eea We can choose for example
$\phi^1_{\rm cl}$ (resp. $\phi^2_{\rm cl}$, $\phi^3_{\rm cl}$) to
be the instanton components of the field $Z$ (resp. $\Phi$,
$\Psi$). To calculate instanton contributions to the BMN
correlators we write each scalar field in the composite BMN
operator as
 \begin{equation}
\phi^i = \phi^i_{\rm cl} + \delta\phi^i, \quad
\phi_i^\dagger =
\phi_{i \,\rm{cl}}^\dagger + \delta\phi_{i}^\dagger,
\end{equation}
 where
$\phi^i_{\rm cl}$, $ \phi_{i \,\rm cl}^\dagger$  are the instanton
background fields \eqref{phiinst}
\begin{equation}
\label{explicit-form} Z_{\rm
cl}\sim \z^0\z^1,\quad \Phi_{\rm cl}\sim \z^0\z^2,\quad \Psi_{\rm
cl }\sim \z^0\z^3,\qquad \bar{Z}_{\rm cl}\sim \z^2\z^3,\quad
\bar{\Phi}_{\rm cl}\sim \z^1\z^3, \quad \bar{\Psi}_{\rm cl}\sim
\z^1\z^2,
\end{equation}
induced by the exact fermion zero modes, and
$\delta\phi^i$, $ \delta\phi_{i}^\dagger$ represent the
fluctuations of scalar fields in the instanton background (they
also include contributions of the lifted fermion zero modes).
Unless all the sixteen exact fermion zero modes are saturated,
instanton corrections to the correlator $\langle \cO_{\rm BMN} (x)
\bar{\cO}_{\rm BMN}(0)\rangle $ vanish. Due to the form
\eq{explicit-form} and since $(\zeta(x))^n =0$ for $n \geq 3$, it
is obvious that $\cO_{\rm BMN}$ and $\bar{\cO}_{\rm BMN}$ must
each provide  eight fermion zero modes $(\z^0)^2 (\z^1)^2 (\z^2)^2
(\z^3)^2$ in order to be able to saturate the sixteen exact
fermion zero modes. Since we need two powers of $\z^0$ and since
$\z^0$ enters only the fields $Z, \Phi, \Psi$, potentially
non-vanishing contributions can come only from
\begin{equation} \label{zero0}
\cO_{\rm BMN} = (Z \Phi \;\mbox{or}\; Z\Psi\; \mbox{or} \;\Phi
\Psi\; \mbox{or} \; ZZ \; \mbox{or} \;\Phi \Phi\; \mbox{or} \;\Phi
\Psi )_{\rm cl} \times \mbox{(2 more $\phi_{\rm cl}^\dagger$)}
\times \mbox{( $\delta\phi$'s and $\delta\phi^\dagger$'s)}.
\end{equation}
Note that $\cO_{\rm BMN}$ cannot contain $\Phi$ and $\bar{\Phi}$,
$\Psi$ and $\bar{\Psi}$ or $Z$ and $\bar{Z}$ simultaneously
(however ``non-holomorphic''  BMN operators which contain $\Phi$
and $\bar{\Psi}$ are allowed). Using the explicit form
\eq{explicit-form}, one can check that \eq{zero0} can never
generate the required combination $(\z^0)^2 (\z^1)^2 (\z^2)^2
(\z^3)^2$. Similarly, one can show that $\bar{\cO}_{\rm BMN}$ can
never give rise to the zero mode structure  $(\z^0)^2 (\z^1)^2
(\z^2)^2 (\z^3)^2$. Hence we conclude that there are no instanton
contributions to the two-point functions of BMN operators.

Next we consider three-point functions. Using the above analysis,
it is clear that to saturate all sixteen zero modes, two of the
operators must provide six zero modes each, and the remaining one
-- four zero modes. For example, consider the correlator 
\begin{equation}
\langle \cO_1(x) \cO_2(y) \cO_3(0) \rangle 
\end{equation} 
where schematically (i.e. discarding the phase factors and the sums) 
\begin{equation} 
\cO_1 := \tr
[Z^{J_1} \Phi \bar{\Psi}] \quad \cO_2 := \tr [Z^{J_2} \bar{\Phi}
\Psi ],\quad \cO_3 :=  \tr [\bar{Z}^{J}], 
\end{equation} 
with $J=J_1+J_2$.
Now it is easy to see that all sixteen fermion zero modes can be
saturated since
\bea &&\cO_1 \ni (\delta Z)^{J_1-1} (Z \Phi
\bar{\Psi})_{\rm cl}
\sim (\z^0)^2  (\z^1)^2 (\z^2)^2, \\
&&\cO_2 \ni (\delta Z)^{J_2-1} (Z \bar{\Phi} \Psi)_{\rm cl}
\sim (\z^0)^2  (\z^1)^2 (\z^3)^2, \\
&&\cO_3 \ni (\delta\bar{Z})^{J-2} (\bar{Z}^2)_{\rm cl} \sim
(\z^2)^2 (\z^3)^2. 
\eea

We conclude with a few remarks.
\begin{enumerate}

\item We saw that our simple argument based on fermion zero mode
counting does not work for three-point functions of
non-holomorphic BMN operators. To see if the instanton
contribution to these correlators vanishes one would require a
more detailed calculation which would have to take into account
the precise form of the BMN operators -- including the phase
factors, the sums and the specific choice of a re-diagonalized
basis.

\item Instanton contributions to three-point functions of
holomorphic BMN operators with scalar impurities vanishes
automatically.

\item We checked that when the BMN phase factors are absent, the
instanton contributions to all three-point functions vanish due to
a precise term by term cancellation. This was expected since in
this case BMN operators are protected, and their two- and
three-point functions do not receive either perturbative or
nonperturbative corrections \cite{seiberg,dzf3}.

\item Finally, instanton corrections to all BMN two-point
functions vanish automatically and this indicates that the
two-point functions are the preferred building blocks of the
pp-wave/SYM dictionary. This is in agreement with discussions in
the recent literature.

\end{enumerate}

\section*{Acknowledgements}
We would like to thank N. Constable, P. Dorey, D. Freedman, J.
Plefka, S. Ross, R. Russo and M. Staudacher for useful discussions
and G.~Arutyunov, E.~Sokatchev and A.~Volovich for comments on the
manuscript. CSC thanks the Department of Physics, National Tsing
Hua University and the National Center of Theoretical Science,
Taiwan for hospitality. We acknowledge grants from the Nuffield
foundation and PPARC.

\end{document}